\documentclass[9pt,twocolumn]{article}

\usepackage[utf8]{inputenc}
\usepackage[english]{babel}
\usepackage{amsmath}
\usepackage{amsfonts}
\usepackage{amssymb}
\usepackage{mathpazo} 
\usepackage[scaled]{helvet}
\usepackage[T1]{fontenc}
\usepackage{url}
\usepackage{verbatim}
\usepackage{float}
\usepackage{caption}
\usepackage{subcaption}

\usepackage{graphicx}
\usepackage{xcolor}
\usepackage{booktabs}
\usepackage{algorithm}
\usepackage[noend]{algpseudocode}
\usepackage{changepage}

\usepackage[left=48pt,%
                right=42pt,%
                top=42pt,%
                bottom=60pt,%
                headheight=15pt,%
                headsep=10pt,%
                letterpaper,twoside]{geometry}%
                
\usepackage[labelfont={bf,sf},%
                labelsep=period,%
                figurename=Fig.,%
                singlelinecheck=off,%
                justification=RaggedRight]{caption}
                
\usepackage[numbers,sort&compress]{natbib}
\title{Efficient second-harmonic generation of a high-energy, femtosecond laser pulse in a lithium triborate (LBO) crystal}

\usepackage{authblk}
\author[1]{C. Aparajit}
\author[1]{Kamalesh Jana}
\author[1]{Amit D. Lad}
\author[1]{Yash M. Ved}
\author[2]{Arnaud Couairon}
\author[1]{G. Ravindra Kumar}

\affil[1]{Tata Institute of Fundamental Research, 1 Homi Bhabha Road, Colaba, Mumbai 400 005, India}
\affil[2]{Centre de Physique Théorique, CNRS, Ecole Polytechnique, Institut Polytechnique de Paris, 91128 Palaiseau, France}
\affil[*]{Corresponding author: aparajit.c@tifr.res.in; and grk@tifr.res.in}

\date{}

\begin{document}

\twocolumn[
  \begin{@twocolumnfalse} 
    
    \maketitle
    
    \begin{abstract}
    We  demonstrate  the highest  efficiency ($\sim$80\%) second harmonic generation (SHG) of  Joule level, 27 femtosecond, high contrast  pulses in a type-I lithium triborate (LBO) crystal. In comparison, potassium dihydrogen phosphate (KDP) gives a maximum efficiency of 26\%.  LBO thus offers high intensity ($>$10$^{19}$ W/cm$^{2}$), ultra-high contrast femtosecond pulses, which have great potential for high energy density science particularly with nanostructured targets as well as  technological applications. 
    \vspace{2mm}
    \end{abstract}
    
  \end{@twocolumnfalse}
]

\section{Introduction}

Intense, femtosecond laser pulse interactions with solids facilitate high energy density science and provide high brightness, ultrafast x-ray and THz sources on the one hand  and energetic pulsed electron, proton, ion and positron sources on the other- all on a table top \cite{drake_2006, Kaw2017}. The development of high-power, femtosecond lasers sparked off  the study  of ultrafast, intense laser-matter interactions in  great detail \cite{gibbon_2013, Mourou}. The big attraction here is the 'instantaneous' excitation of solid density matter \cite{MURNANE531}, unlike in studies with longer pulses where the  excitation occurs over a partially ionized, expanding, lower than solid density plasma. In practice, an amplified femtosecond pulse has a longer picosecond wing and nanosecond pedestal and replica. These 'prepulses'  are intense enough to create a significant density preplasma on a target, particularly at the highest peak intensities (10$^{22}$ W/cm$^{2}$) \cite{Bahk:04}, which can adversely influence the interactions between the subsequent, main laser pulse and the solid.  The key parameter therefore is the intensity contrast, i.e. the ratio of the peak intensity of the femtosecond part  to that at the picosecond  pedestal level. Much effort has therefore been made for cleaning up prepulses \cite{Chuang:91, Nantel}. \\
 
  A major parallel direction in the recent past has been the optimization of structured targets for producing hot, dense plasma conditions and hence  high flux, electromagnetic and material particle emissions \cite{drake_2006, Kaw2017}. Some of the well known structures include nanoparticles \cite{Rajeev-PRL2003}, nanowires \cite{Purvis2013, Samsanova}, nanobrushes \cite{Zhao_Nanobrush} and   gratings \cite{Macchi, Kahaly-PRL2008}.  The necessity of a clean femtosecond pulse becomes particularly acute here because the prepulses can easily destroy the surface structures before arrival of the main pulse and hence frustrate its very purpose \cite{Gibbon_2007}.  The  enhancement of the local electric fields by these structures works against their survival at the intensities in the picosecond pedestal thereby causing enhanced preplasma. \\
  
  The picosecond pedestal is suppressed by a fast saturable absorber\cite{Chuang:91}, cross-wave polarization gating \cite{Jullien:05}, the seeding of the amplifiers  at high energy and improving compressor components \cite{Hooker:11, Hooker2}.  The first two involve nonlinear optical processes. In lasers systems with intrinsic low contrast, external plasma mirroring \cite{Thaury2007} and harmonic generation \cite{Hillier:13} can improve contrast. Second harmonic generation (SHG) is particularly attractive, since it is simple, robust and easy to control.  In practical terms, the application of SHG is dependent on  the availability of highly efficient, large crystals (to accommodate large amplified beams) with  manageable dispersion and high damage threshold. Given these several constraints, there have been very few studies on this topic, particularly at high intensities \cite{Hillier:13, Hornung, Aoyama:01, Mori, Marcinkevicius}. \\

  In this paper, we demonstrate  SHG of 800 nm, high contrast (10$^ {9}$), 27 femtosecond  pulses at nearly 80\%  efficiency in a lithium triborate (LBO) crystal at peak incident intensities as large as 1400 GW/cm$^{2}$. The resulting 400 nm, ultra-high contrast pulses can generate focused intensities as large as 5 x 10$^{19}$ W/cm$^2$ (considering focal waist to be 10 $\mu$m).  We present variation of efficiency with input laser energy and the corresponding SHG spectra.  We compare LBO  with potassium dihydrogen phosphate (KDP)  under the same conditions and compare our results in context of other studies.\\

\section{Experiment}

\begin{figure}[!ht]
    \centering
    \includegraphics[width=0.95\linewidth]{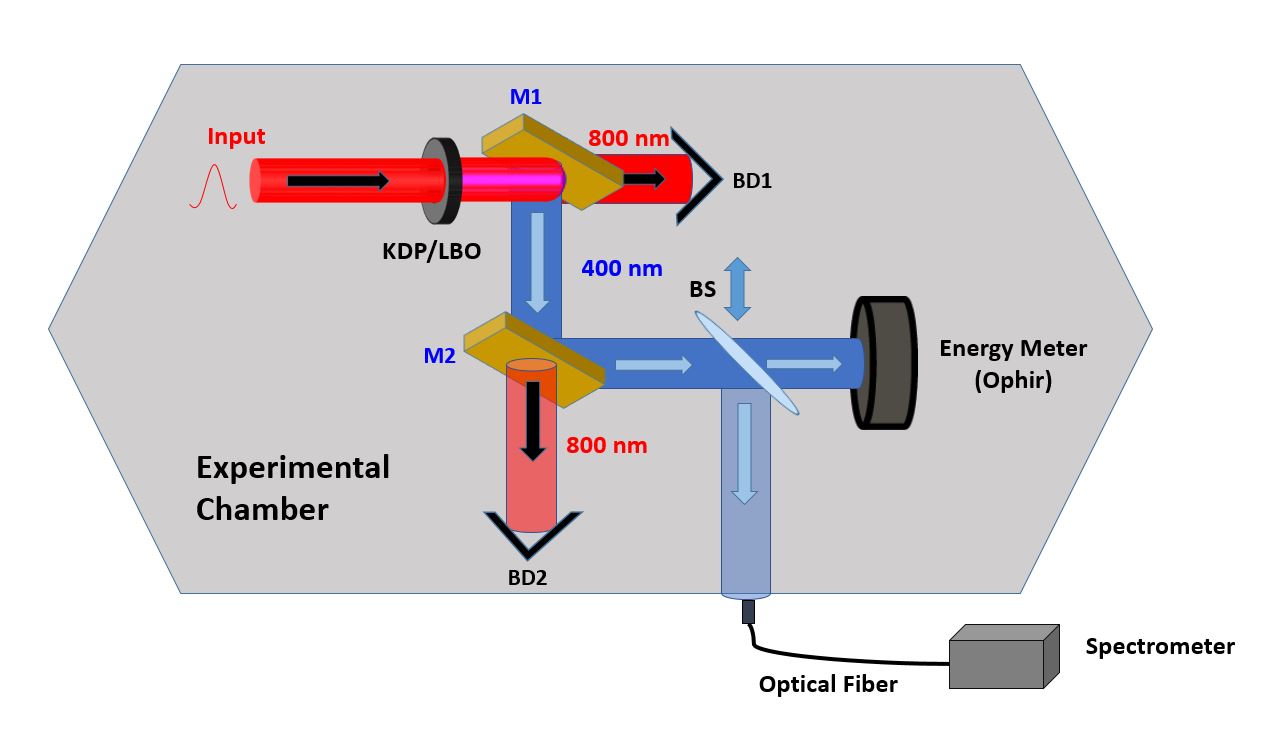}
    \caption{Schematic of the experimental setup (M1-M2: Dichroic Mirrors for 400 nm, BS: Beamsplitter, BD1-BD2: Beam Dumps).}
    \label{fig:setup}
\end{figure}

\begin{figure}
    \centering
    \includegraphics[width=\linewidth]{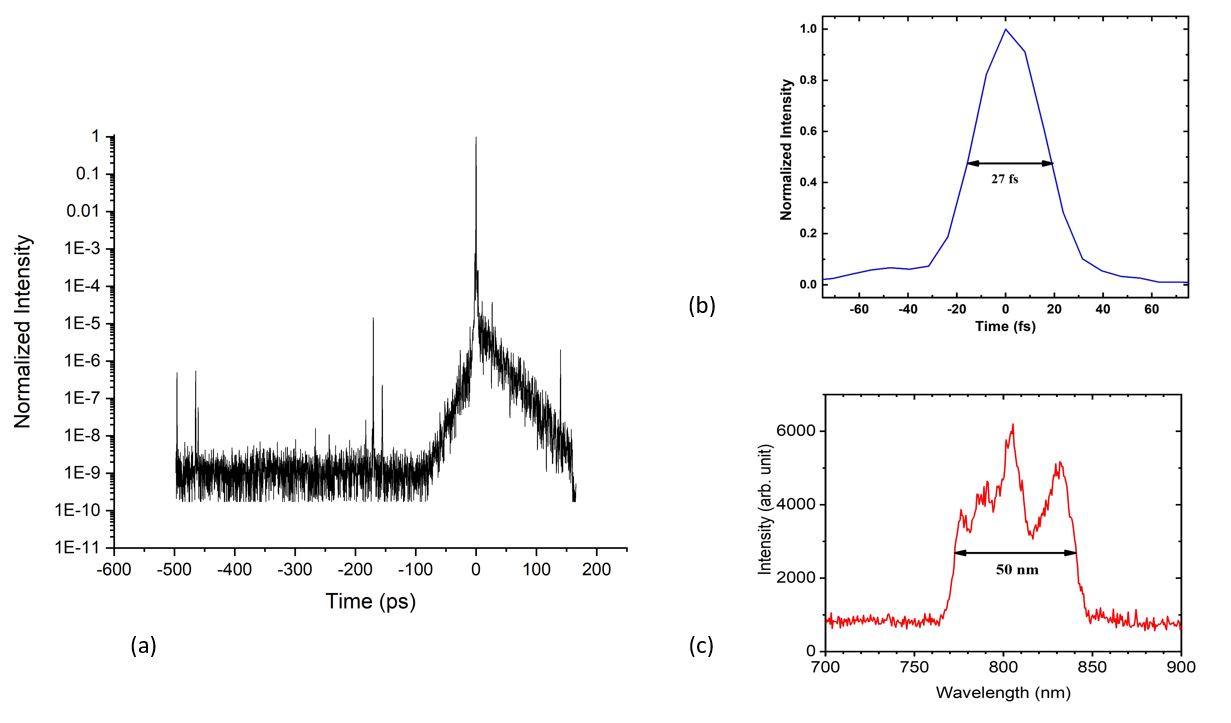}
    \caption{(a) Intensity contrast (measured using SEQUIOA), (b) pulse-width (measured using SPIDER), and (c) bandwidth, of fundamental 800 nm pulse. SPIDER - Spectral phase interferometry for direct electric field reconstruction (SPIDER) \cite{Iaconis:98} (Amplitude Technologies); SEQUIOA (Amplitude Technologies) - a third-order cross correlator \cite{Eckardt, ALBRECHT198159}. \\
 }
    \label{fig:inputchar}
\end{figure}

 The experiment (Fig. \ref{fig:setup}) was conducted  using a  chirped pulse amplified 150 TW Ti: sapphire laser system delivering 800 nm, 27 fs, 50 nm bandwidth pulses at a 5 Hz repetition rate (Fig \ref{fig:inputchar}).\\
 
The complete experiment was done inside a vacuum chamber (10$^{-5}$ mbar pressure). A type-I lithium triborate (LBO) crystal (Cristal Laser; 70 mm clear aperture, 2 mm thick, anti-reflection coated for both 800 nm and 400 nm) was used for SHG. Two dichroic mirrors coated for 400 nm (R > 99.9\%) were placed after the LBO to eliminate residual fundamental radiation (800 nm). A pyroelectric detector (OPHIR PE100BF-DIF-C) was placed after the two mirrors to measure the SHG energy. After the energy measurements, a beam splitter was placed, and a small fraction (1\%) of the second harmonic (SH) beam was sent through a fused silica window of the vacuum chamber, in order to measure the spectrum of  the SH with a single shot spectrometer (Avantes UV/VIS/NIR AvaSpec-ULS2048L-EVO (200-1100 nm). Similar measurements were obtained for a type-I KDP crystal (52 mm aperture and 1 mm thick; no anti-reflection coating). Crystal characteristics are presented in Table \ref{tab:crystal2}. \\

\begin{table}[!ht]
    \centering
    \begin{tabular}{ |c||c|c| } 
         \hline
          & LBO & KDP \\ 
         \hline
         Type & Type-I at 800 nm & Type-I at 800 nm \\
         $\theta$ & 90$^{\circ}$ & 44.9$^{\circ}$ \\ 
         $\phi$ & 31.6$^{\circ}$ & 45$^{\circ}$ \\ 
         Diameter & 70 mm & 52 mm\\
         Thickness & 2 mm & 1 mm\\
         Manufacturers & Cristal Laser & EKSMA Optics \\
         
         \hline
    \end{tabular}
    \caption{Charateristics and dimensions of the SHG crystals.}
    \label{tab:crystal2}
\end{table}

\begin{figure*}[!ht]
    \begin{subfigure}{.45\linewidth}
        \centering
		\includegraphics[width = 0.95\linewidth]{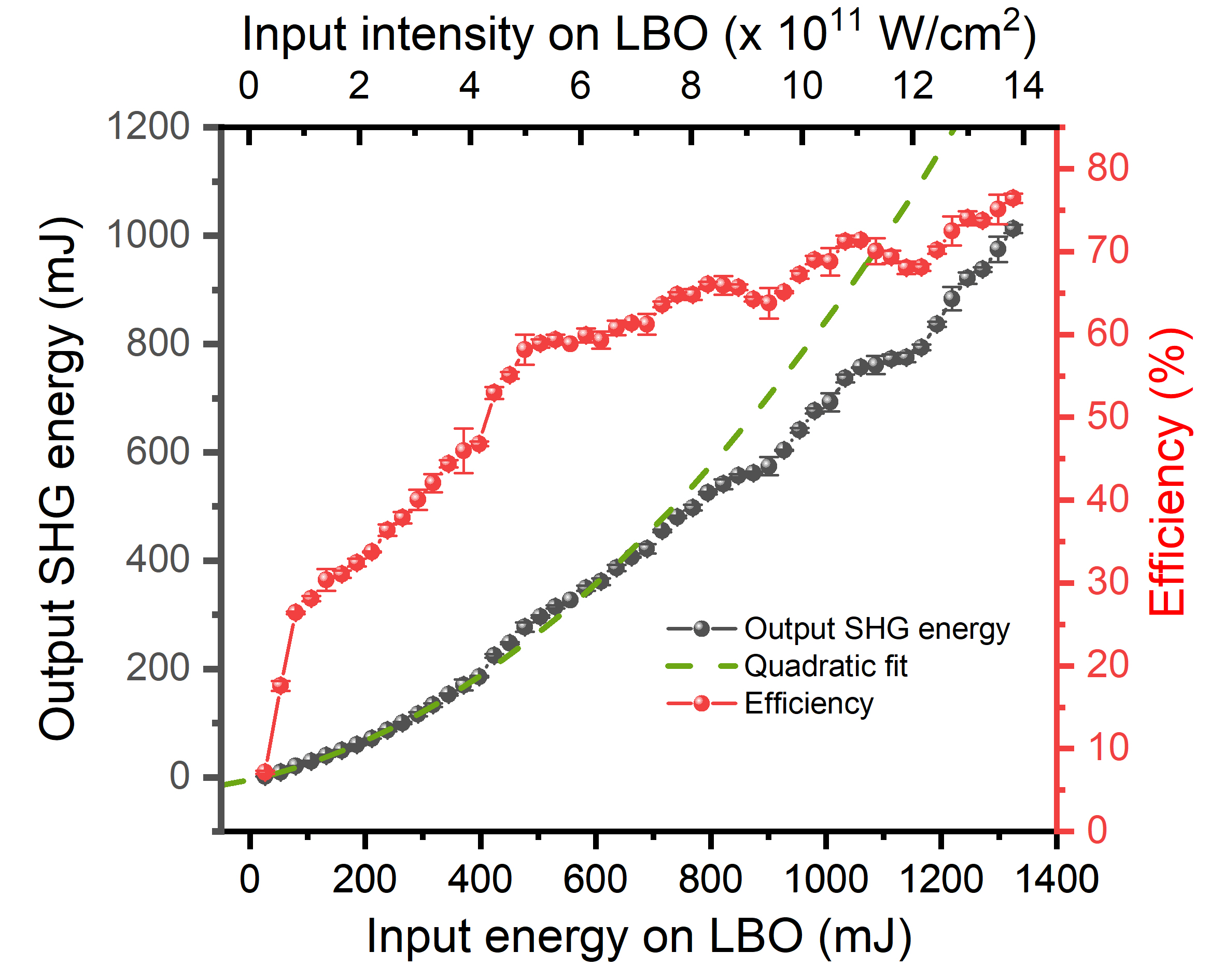}
		\caption{}
		\label{fig:eff1}
    \end{subfigure}
    \begin{subfigure}{.45\linewidth}
        \centering
        \includegraphics[width = 0.95\linewidth]{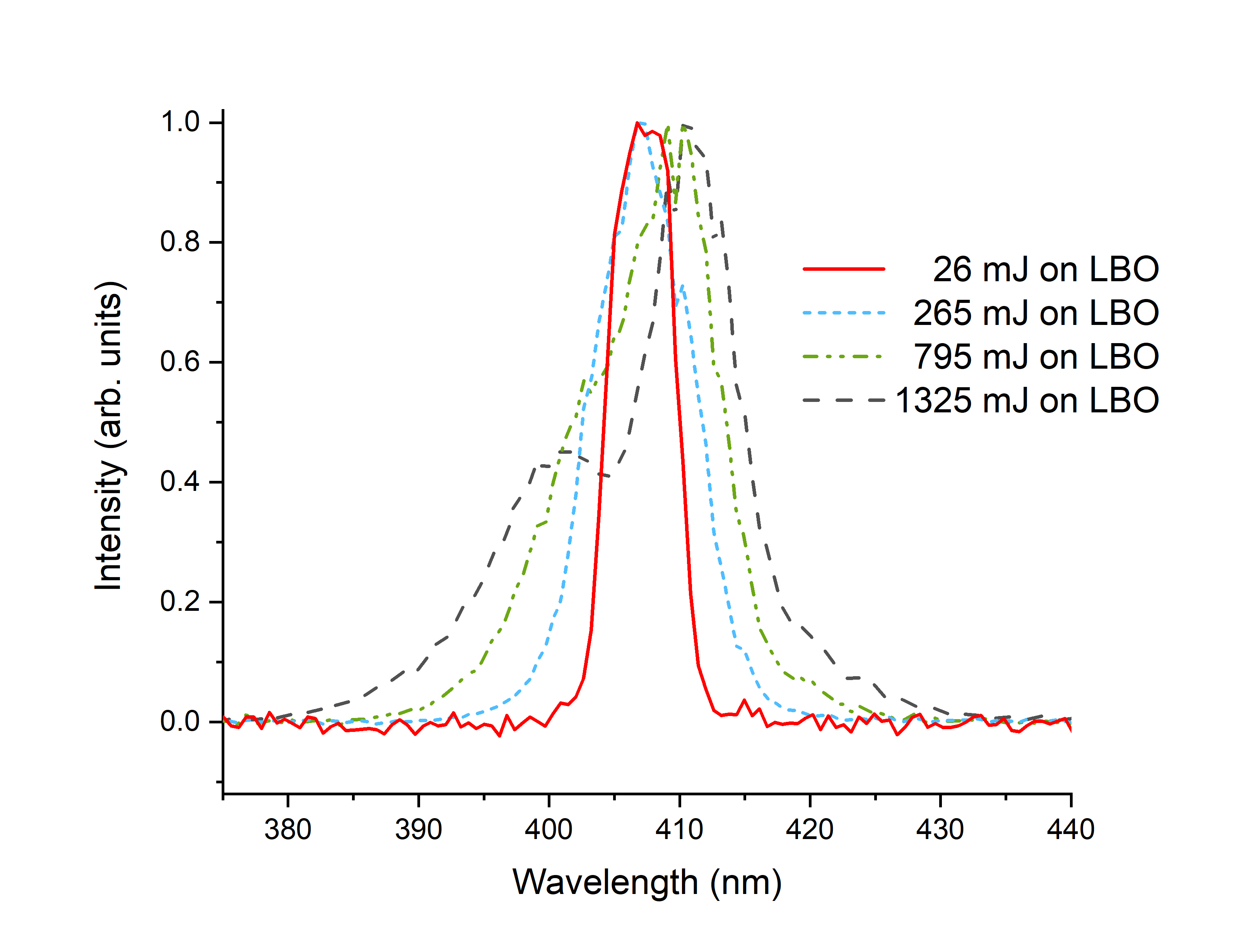}
        \caption{}
        \label{fig:lbospec}
    \end{subfigure}
    
    
    \begin{subfigure}{.45\linewidth}
        \centering
		\includegraphics[width = 0.95\linewidth]{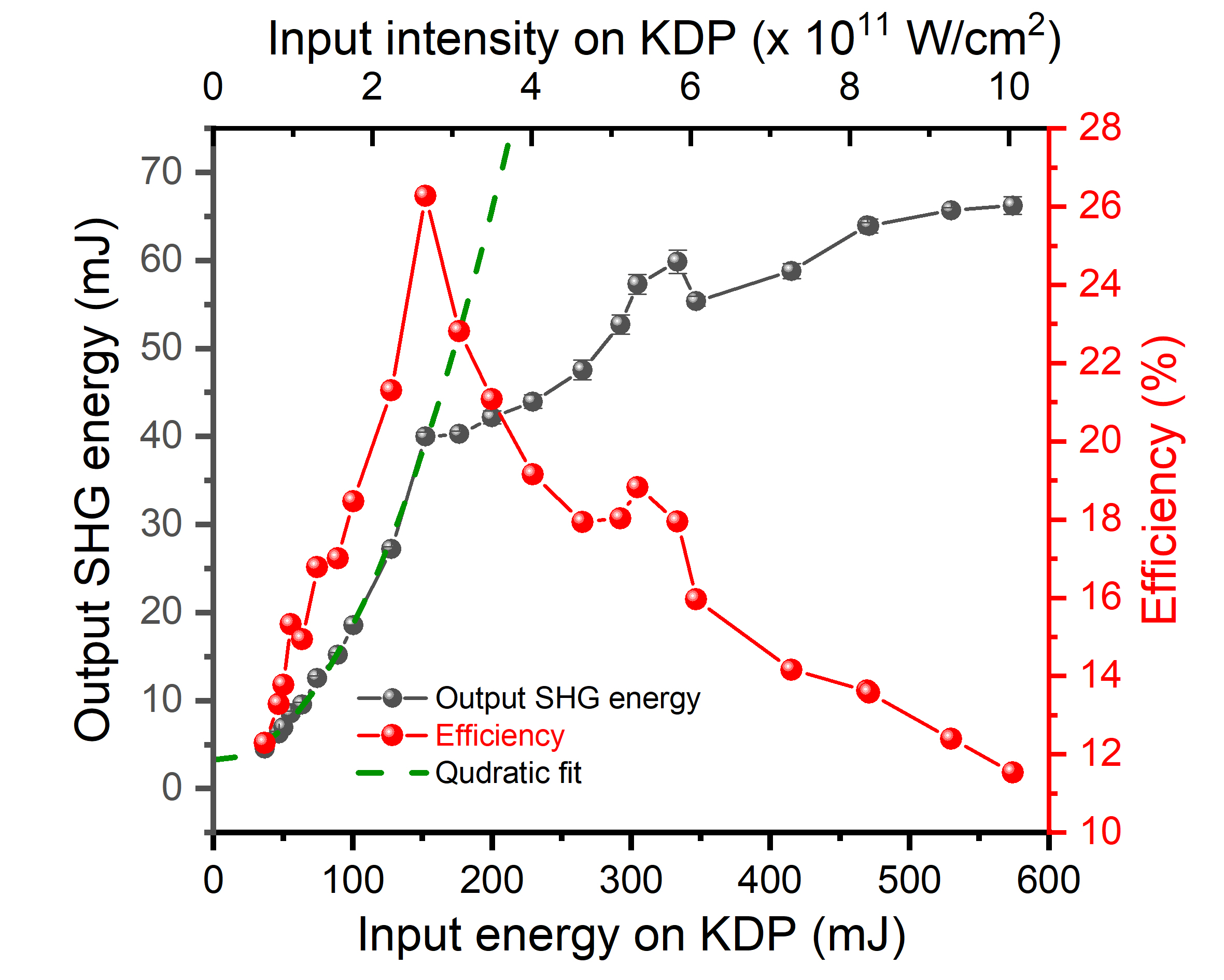}
		\caption{}
		\label{fig:kdp1}
    \end{subfigure}
    \begin{subfigure}{.45\linewidth}
         \centering
         \includegraphics[width = 0.95\linewidth]{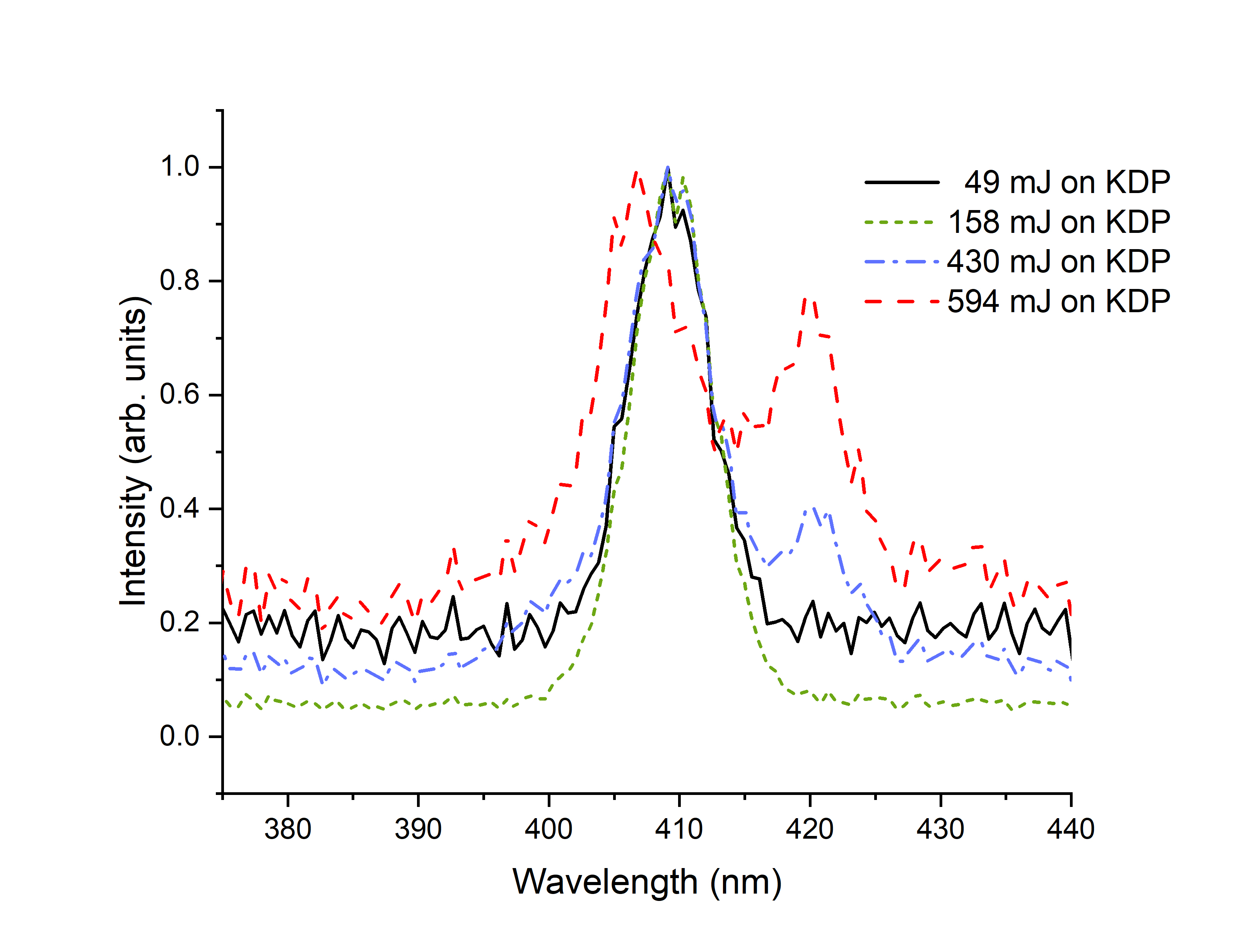}
         \caption{}
         \label{fig:kdpspec}
    \end{subfigure}
    
    \caption{(a) SHG Conversion efficiency from LBO crystal, (b) SHG spectrum of LBO crystal by varying input laser energy, (c) SHG Conversion efficiency from KDP crystal, (d) SHG spectrum of KDP crystal by varying input laser energy. }
    \label{fig:fig2}
\end{figure*}

 \begin{table*}[!ht]
    \centering
    \begin{tabular}{ |p{4cm}||c|c|c|c|c|c| } 
         \hline
          Parameters & D. Hillier & M. Hornung & M. Aoyama & K. Mori & A. Marcinkevičius & Ours \\ 
         \hline
         $\lambda$ (nm) & 1054 & 1030 & 800 & 800 & 800 & 800\\
         Pulsewidth (fs) & 500 & 143 & 100 & 125 & 130, 180 & 27\\ 
         Bandwidth (nm) & 3 & 10 & 10 & 9 & 8, 6 & 50\\ 
         Intensity (GW/cm$^2$) & 250 & 487 & 192 & 100 & N/A & 1100, 1400\\ 
         Crystal used & KDP & KDP & KDP & KDP & KDP & KDP, LBO\\ 
         Crystal diameter (mm) & 320 & 160 & N/A & N/A & 65 & 52, 70\\
         Crystal thickness (mm) & 3 & 2 & 1 & 1 & 2 & 1, 2\\
         Maximum Efficiency (\%) & 75 & 30 & 80 & 60 & 45, 45 & 26, 80 \\ 
         
         \hline
    \end{tabular}
    \caption{Comparison of SHG conversion from different experiments. D. Hiilier et. al. \cite{Hillier:13}, M. Hornung et. al. \cite{Hornung}, M. Aoyama et. al. \cite{Aoyama:01}, K. Mori et. al. \cite{Mori}, A. Marcinkevičius et. al. \cite{Marcinkevicius}; N/A - not available.}
    \label{tab:crystal}
    
\end{table*}

\section{Results and Discussion}

Fig. \ref{fig:eff1} shows the SHG energy and conversion efficiency as a function of both input energy and input intensity on the LBO. The error bars arise from repeated measurements. At lower intensities, SHG process has an output that scales as the square of the input energy, as seen from the quadratic fit.  At high intensities used in the experiment however, the output energy is seen to deviate from the quadratic behaviour into saturation. The SHG output of KDP (Fig. \ref{fig:kdp1}) similarly shows an initial quadratic increase with input, reaching a maximum of around 26\% at 150 mJ (or an intensity of $\sim$250 GW/cm$^2$) before saturation of the output SHG energy. At high intensities, saturation is expected from significant coupling to higher order nonlinearities \cite{Ditmire:96}. In particular, nonlinear spatial and temporal modulation  of the input and generated pulses can lead to significant group velocity mismatch and walk- off between the fundamental and the second-harmonic (SH) pulses. The efficiency  reaches  an attractive 50\%   at  energies as low as 400 mJ and a peak of about 80\% at 1.2 J. We believe such conversion efficiencies using high-intensity, sub-100 femtosecond pulses have not been measured in other studies and hence are really promising for the field of high energy density physics, in particular ultrahigh contrast experiments at relativistic intensities. For example, the  generated 400 nm pulse has been focused to intensities of 10$^{19}$ W/cm$^2$ in experiments  currently underway in our laboratory. \\

Fig. \ref{fig:lbospec} shows the spectrum of SH pulses for different input energies on the LBO. We note that the bandwidth of the spectra increases with input energy and develops additional modulations. Such modulations are known to arise from higher order nonlinearities like self phase modulation \cite{Ditmire:96}. At lower input energy  the bandwidth is about 6 nm and at higher input energy, about 9 nm. The higher bandwidth implies that the SH pulse generated at higher intensities could potentially have a lower pulse-width, with respect to Fourier transform limit arguments ($\Delta \nu \Delta \tau = k$). But, the true pulse width of the SH generated pulse is hard to estimate due to a variety of effects related to walk-off and higher-order nonlinear effects. Fig. \ref{fig:kdpspec} shows the SH spectra for different input intensities from the KDP crystal. It also similarly shows that for higher intensities the SH spectra are broadened and modulated. \\

Comparing the results for SHG efficiencies using KDP and LBO, we see much larger efficiencies for the LBO. The difference in crystal thickness plays a role between the two results by a factor of four. Also, previous studies reveal LBO to have superior damage threshold, higher nonlinearity, smaller walk-off angle and larger acceptance angle as compared to KDP \cite{Nikogosyan1994, Lin_LBO}, which could explain the differences in efficiencies seen. Another reason for the difference could be that in the case of KDP, conversion into higher order nonlinearities starts at lower intensities than for LBO. Also, lower damage threshold of the KDP crystal makes it difficult to  use higher intensities on them. \\

Let us compare our results with earlier reports of SHG (Table \ref{tab:crystal}). Firstly, we have used the highest intensities on the crystal, possible for femtosecond pulses, and also our pulse bandwidth is the highest. Hence, we face the most adverse effects of walk-off and possible interference of higher order nonlinearities, given that our intensities have reached $~$ 1400 GW/cm$^{2}$. Nevertheless, it is gratifying to note that we have reached conversion efficiencies as large as 80\%, which is extremely encouraging for applications particularly in high energy density science where the whole purpose of the SHG process is to improve the intensity contrast of the pulses. We are able to achieve intensities as large as 5 x 10$^{19}$ W/cm$^{2}$ in experiments that are ongoing. In comparison with other results however, it is to be noted that the largest efficiencies reported were by Hillier et. al. at 75\%  \cite{Hillier:13} and Aoyama et. al at 80\% \cite{Aoyama:01}. Hillier et. al used 500 fs pulses at much larger input energies, but also much larger crystal sizes, keeping the intensity on the crystal rather low (250 GW/cm$^{2}$). Note that their pulses had a bandwidth of only about 3 nm and thus much less walk-off effects, due to which they were able to use larger crystal lengths. Similarly, Aoyama et. al used 100 fs pulses with a bandwidth of about 10 nm, and intensity on crystal around 192 GW/cm$^{2}$ - compared to our experiment, longer pulsewidth and shorter bandwidth.  \\

Secondly, at high intensities in our case, we do see the effects of pulse spectral broadening which is indicative of higher order nonlinear effects \cite{Ditmire:96}. However that doesn't seem to affect the conversion efficiency. It may however have implications for the SHG pulse duration. Preliminary measurements that we have made reveal a pulse duration of about 50 fs. Also, we are in the process of setting up a self-diffration FROG \cite{Trebino-Kane} at 400 nm which should give us the true measurement of the pulse duration and instantaneous phase, which we shall report in future publications.   \\

Finally, the contrast of the SHG pulse is not measurable with the current technology. We can however estimate  the best value to be the square of the contrast of the input pulse,  the latter being 10$^{9}$ in our case. In practice, it may not be exactly the best achievable but we can estimate an improvement in the picosecond contrast to be at least a million fold \cite{Hillier:13}.

\section{Conclusions}

We report a highly attractive $~$80\% SHG efficiency from a 2 mm thick LBO crystal, using 800 nm, 27 femtosecond and 50 nm bandwidth, 10$^9$ intensity contrast pulses. In comparison, we get only $~$26\% from a 1 mm KDP crystal. Efficient SHG from such high bandwidth pulses requires proper consideration of effects due to walk-off, which adds more appeal to this result. We also show SHG spectral broadening at high intensities, indicating  the effects of higher order nonlinearities.  Due to such high efficiencies, we are able to reach intensities $>$ 10$^{19}$ W/cm$^2$ when focused on a target. In addition, SHG can improve the intensity contrast by many orders of magnitude, hence these pulses can facilitate clean femtosecond experiments with structured targets to produce hot, dense plasma conditions and ultrabright radiation sources. \\

\noindent\begin{Large}
\textbf{Funding}.\end{Large} GRK acknowledges a J.C. Bose Fellowship grant (JCB-037/2010) from the Science and Engineering Board (SERB), Government of India. \\

\noindent\begin{Large}
\textbf{Acknowledgement}. \end{Large}The authors thank K. Chandra Vardhan for help in the experiment. \\


\bibliography{references}
\bibliographystyle{ieeetr}


\end{document}